\documentstyle[preprint,aps]{revtex}
\font \afont = cmmi10 scaled \magstep 2
\mathchardef \A="0861 \rm
\mathchardef \B="0862 \rm
\begin{document}
\draft
\preprint{QCFT-3.3-12/17/95}
\title{Quantum-conformal Field Theory}
\author{Daniel C. Galehouse}
\address{Physics Department, University of Akron, Akron, Ohio 44325}
\date{\today}
\maketitle
\begin{abstract}
A theoretical study is made of conformal factors in certain types of
physical theories based on classical differential geometry.  Analysis of
quantum versions of Weyl's theory suggest that similar
field equations should be available in four, five and more
dimensions. Various conformal factors are associated with the
wave functions of source and test particles.
This allows for certain quantum field equations to be developed.
The curvature tensors are calculated and separated into gravitational,
electromagnetic and quantum components.  Both four and five dimensional
covariant theories are studied.
Nullity of the invariant five scalar of curvature leads to the Klein-Gordon
equation.
The mass is associated with an eigenvalue of the differential operator
of the fifth dimension.    Different concepts of interaction are possible
and may apply in a quantum gravitational theory.
\end{abstract}
\vfill \eject
\narrowtext
\section{INTRODUCTION}
\label{int}

This article is the second in a sequence  that discusses how
differential geometry is capable of describing quantum
mechanical effects.  The objective is to obtain compatible  quantum
field equations that complement the geodesics
of the preceding article.  Important results for a single particle
field equation are presented. Results for the a complete set of source
equations is not contained herein but some of the
initial problems are addressed.
The usual methods of developing field equations are not effective
for this type of construction and are counterindicated by the considerations
of~\cite{r18}.   The standard methods fail and more
elementary arguments are used to guide the
developments.  Progress  is slow because
the basic equations must be found by physical
principles, such as equivalence, or by
the mathematical study of limiting forms of the curvature tensor.
It is a process of discovery in elementary geometrical systems.

The issue of the particle mass is crucial and
some of the characteristic roles are resolved
in a synthesis that includes quantum effects.  The mass as a constant
must be introduced in a way that is compatible with the characteristic
properties of all fields.
One would wish to develop, as in the Weyl theory~\cite{r20,r19,r51,r53},
a Klein-Gordon equation~\cite{r1,r2} in some form.
Some guidance is available as there are various theories in the
literature~\cite{r3,r3a,r4,r5,r6,r7,r8,r9a,r10,r56}
that apply in an approximation or in a limiting case.

Particle motion as discussed in the preceding paper~\cite{0r}
"Quantum Geodesics", (QG) has not been assigned a particular particle mass.
This defect is to be remedied here. Five dimensional
field equations  are developed in which the
mass appears as a measure of inertia.  An appropriate construction is
suggested by a version of the Weyl theory.
An identified relationship with quantum phenomenology has  been known
 for some time.
These associations provide a way to identify quantum structures
in other types of geometry.   To do this, the Weyl theory is
reviewed and rewritten in a form that is appropriate.   This is applied to
Riemannian theories and in five dimensions, the field equations can be
combined with the geodesics of the previous article.

\section{QUANTUM EXTENSION OF WEYL GEOMETRY}
\label{qwg}

Although the mathematics of Weyl's theory
has been known since 1918, the physical interpretation has been in dispute.
Any non-Riemannian geometry can be interpreted as a modification
to the derivatives as they are applied microscopically to particle motion.
The usual relation between the coefficients of connection and the
Christoffel symbols  is changed.
An arbitrary tensor $D^\beta_{\mu\nu}$ is  added to give a new
connection~\cite{r19}.
\begin{equation}
\Gamma^\beta_{\mu\nu}={\beta \brace \mu\nu}-D^\beta_{\mu\nu} \label{nrcc}
\end{equation}
The tensor  $D^\beta_{\mu \nu}$ is chosen symmetrical
in $(\mu \gamma)$.  Infinitesimal displacements, calculated from
these new values $\Gamma^\beta_{\mu\gamma}$ may now cause a
change in the length of a displaced vector besides the usual
change in direction.  Previous studies show that quantum and
electromagnetic effects are correctly included if
\begin{equation}
D^\beta_{\mu\nu}=
\delta^\beta_\mu(\phi_\nu - \ln\psi|_\nu)
+\delta^\beta_\nu(\phi_\mu - \ln\psi|_\mu)
-g_{\mu\nu}(\phi^\beta - \ln\psi|^\beta) \label{wegu}
\end{equation}
where $\psi$ is the wave function and $\phi_\mu=ieA_\mu$.
The  effect of a displacement around a
closed loop can be used to define a more general Riemann tensor.
\begin{equation}
\delta V^\mu=R^\mu_{\nu\lambda\beta}V^\nu \delta x^\lambda x^\beta
\end{equation}
with the definition
\begin{equation}
R^\mu_{\nu\lambda\beta}=
{\partial \Gamma^\mu_{\nu\beta} \over \partial x^\lambda}-
{\partial \Gamma^\mu_{\nu\lambda} \over \partial x^\beta}+
\Gamma^\gamma_{\nu\beta}\Gamma^\mu_{\gamma\lambda}-
\Gamma^\gamma_{\nu\lambda}\Gamma^\mu_{\gamma\beta}. \label{wrcc}
\end{equation}

With the implied complex factors in the connections given
buy~(\ref{nrcc}) and~(\ref{wegu}), the
usual non-integrable part of the Weyl displacement
is associated with the phase.  The part of Weyl's connections which
would affect the scale of $g_{\mu\nu}$ are integrable.
For this real conformal factor, and
in this gauge, Einsteins's objection to Weyl's theory does not hold
{}~\cite{r20,r53}.  The integrable part of the length change
seems to be mathematically
related to the gravitational red shift, although the quantum interpretation
is not equivalent to the original Weyl interpretation.

There are a number of theories that study invariance
under  conformal transformations~\cite{r54,r55,r9}.   This
point of view may apply in the classical limit.
The  quantum geometrical system used here departs from this
interpretation and supports well defined
physical meanings for microscopic
conformal transformations.  Invariance under conformal transformations,
especially in classical theories,
is identified as a means to suppress the intrinsic quantum terms.

Conformal transformations of the four metric can be identified with
variations of probability density.  This applies to single particle states.
Consider an observer's space that, for the argument, is flat and
euclidean.  As represented in the diagram, the probability density
is measured by counting the number $N$
of particles in a small region of dimensions $(\Delta x,\Delta y,\Delta z)$
having volume $\Delta V$.  The predicted
probability density, $P^0$, is taken as the first component of the quantum
conserved current density.
Let the particle be described by the metric $g_{\mu\nu}$,
which varies from the observers' metric $\dot g_{\mu\nu}$ by a point
conformal transformation, $g_{\mu\nu}=\lambda \dot g_{\mu\nu}$.
The number of particles in the region, as determined by the
neutral observer is given by
\begin{equation}
{N=P_0\Delta x \Delta y \Delta z \sim \psi^\ast
{\partial \over \partial t} \psi \Delta x \Delta y \Delta z }
\end{equation}
Of course the geodesics of $g_{\mu\nu}$ are not straight in the
observers frame, but
in a very small region where both coordinate systems are flat,
they can be chosen euclidean and parallel.
The fixed volume $(\Delta x,\Delta y,\Delta z,dt)$ has new dimensions
$(\Delta x^\prime,\Delta y^\prime,\Delta z^\prime , dt^\prime)=
(\Delta x,\Delta y,\Delta z,dt)\lambda^{(1/2)}$
while the number of counts stays constant.

The time dependence of the derivative under  conformal transformation
must be compensated
in the same way as the spatial coordinates.  The implied relation
between $g_{\mu\nu}$ and $\psi$ during a gauge transformation can be used
to show that the number of counts in a fixed region is constant.
Transforming
\begin{equation}
{g_{\mu\nu} \to g_{\mu\nu}^\prime=\lambda^\prime \dot g_{\mu\nu}}
\end{equation}
\begin{equation}
{\psi \to \psi^\prime={\psi \over \sqrt{\lambda^\prime}}}
\end{equation}
it follows that
\begin{equation}
{N \to N^\prime=\psi^{\ast\prime}
{\partial \over \partial t^\prime}\psi^\prime
\Delta x^\prime \Delta y^\prime \Delta z^\prime=
\psi^\ast{\partial \over \partial t}\psi
\Delta x \Delta y \Delta z }
\end{equation}
Of special interest is that if for a given solution $\psi$, the
value of $\lambda$ is chosen equal
to the value of $|\psi|^2$ then after the gauge of
$g_{\mu\nu}= \dot g_{\mu\nu}$ is changed to
$g_{\mu\nu}= \lambda \dot g_{\mu\nu}$, the magnitude of the wave function is
forced to a constant over the whole space.
Probability density information can be transferred to the metric.
The observable quantities are invariant although the magnitude of
the wave function has been transformed away and is no longer
an independent field.

\section{QUANTUM-GEOMETRIC FIELD EQUATIONS}
\label{qge}

The field equations of quantum mechanics can be made part of this
geometrical structure.  It is easiest to first consider a four
dimensional example that ignores electrodynamics.
For this relatively simple geometry,
the Klein-Gordon equation can be identified with a
modified type of curvature scalar.  To do this, geometrical invariants
may be formed from the two metrics and derived tensors.
An appropriate scalar is formed by using the tensor $R^\beta_{\mu\lambda\nu}$
calculated from $g_{\mu\nu}$ and contracting it with the inverse of the
observer's metric, $\dot g^{\mu\nu}$.

This imposes a external length standard on the quantum
equations that is based on the observer's frame.  It also implies that
phenomenological measurements performed to determine the observer's
metric, $\dot g_{\mu\nu}$ must be made in a way that is consistent
with quantum mechanics.  In particular, the clocks must all be based
on fundamental
quantum processes.
Direct calculation gives
\begin{equation}
R^\mu_{\nu\mu\lambda} \dot g^{\lambda \nu}=6{\Box \psi \over \psi}
\end{equation}
where $g_{\mu\nu}=\psi^2 \dot g_{\mu\nu}$ with $\dot g_{\mu\nu}$
still assumed euclidean.  This means
that the identification
\begin{equation}
R^\mu_{\nu\mu\lambda} \dot g^{\nu\lambda}=-6m^2 \label{winkg}
\end{equation}
is equivalent  to the Klein-Gordon equation
\begin{equation}
 \Box \psi=-m^2 \psi.
\end{equation}
The field equation for $\psi$ is demonstrated to be part of the
geometry along with the probabilistic interpretation.

Thus, single particle relativistic quantum mechanics, can be described
by the distortions of a kind of space-time.  A larger system is needed to
include a complex wave function and practical interactions.
In particular the five dimensional theory,
helps resolve some of the problems.  To develop a usefulinterpretation,
the Weyl theory must be rewritten without an explicit wave function.
To do this, the gauges must be adjusted.
Starting with some combination $\psi$, $A_\mu$ and
$g_{\mu\nu}$, the gauge is transformed
so that $\psi$ is equal to unity everywhere.  At the
same time both $A_\mu$ and $g_{\mu\nu}$ are kept real.
This leaves the metric unique up
to a constant multiplicative factor and the vector potential
completely specified.  These new field variables, $g_{\mu\nu}$ and
$A_\mu$ are applicable to other types of geometrical theories.

In the case where $A_\mu$ is integrable, the corresponding congruences
are mathematically geodesics.  In this situation $\phi_\mu$
can be combined into a complex Christoffel symbol.  Notwithstanding
the interpretational problems, the trajectories are geodesics for a
complex path parameter.
The difficulties of defining geodetic motion for
non-integrable $A_\mu$ are relieved in the five dimensional theories.
The  significant mathematical quantities are more accessible.
The imaginary parts that naturally appear in the Weyl theory
are often related to the variation with proper time rather
than with coordinate time.  Because the fifth signature element in the
five metric is of sign opposite to that for the coordinate time, this
dependency can appear with the factor $i$.  The problems of changing the
integrability of the vector potential is not easily resolved in a
Weyl theory.

The additional transformations involving $\tau$ must be considered
point transformations that may not leave physical quantities invariant.
The effect on a Weyl connection can be calculated in a way that is similar
to the construction in~\cite{r18}.
If it is applied to the four dimensional neutral space
connections, the Weyl connection is generated.
The electromagnetic term in the Weyl connection can be thought of
as generated by derivatives with respect to $\tau$ which persist after
a series of conformal and cut transformations.  The terms in $\A_\mu$
are multiplied by the quantity $im$ which produces the imaginary factor.
The $\tau$ derivative must be allowed to annihilate on an exponential term.
The coefficients in $\A_\mu$ are multiplied by the quantity $im$.
If a conformal transformation is included,
an electromagnetic field is generated.
The Weyl theory can be thought of as a study of conformal waves in which
the five dimensional coordinate
transformations generate apparent non-Riemannian effects.
In this context, the Weyl theories are a phenomenological expression
of higher geometries.

Calculation of the analogous invariant
$R^\mu_{\nu\gamma\beta} \dot g^{\nu\beta}$ using the full
connection~(\ref{nrcc}) includes
additional terms in the vector potential and metric.
These reduce to the usual Klein-Gordon equation as shown in \cite{r19}.
These identified quantum-Weyl field equations can be rewritten
by recalculating from the invariant~(\ref{winkg}).
Keeping $\lambda$ explicit, the contraction of the metric tensor
is first written:
\begin{equation}
R^\nu_{\mu\nu\beta} \dot g^{\mu\beta}
=\lambda R^\nu_{\mu\nu\beta} g^{\mu\beta}
\end{equation}

Calculation of the ``pure'' invariant $R^\nu_{\mu\nu\beta} g^{\mu\beta}$
follows from equation~(\ref{wrcc}),~(\ref{nrcc}), and ~(\ref{wegu}).
The Weyl curvature can be computed as
\begin{equation}
R^\beta_{\cdot \mu\nu\rho}=Q^\beta_{\cdot \mu\nu\rho}
-D^\beta_{\mu\nu}|_\rho+D^\beta_{\mu\rho}|\nu
-D^\beta_{\rho\tau}D^\tau_{\mu\nu}+D^\beta_{\nu\tau}D^\tau_{\mu\rho}
\end{equation}
where
\begin{equation}
Q^\beta_{\cdot \mu\nu\rho}=
{\partial \over \partial x^\rho}{\beta \brace \mu\nu}
-{\partial \over \partial x^\nu}{\beta \brace \mu\rho}
+{\beta \brace \tau\rho}{\tau \brace \nu\mu}
-{\beta \brace \tau\nu}{\tau \brace \rho\mu}
\end{equation}
is the Riemannian tensor of the undotted metric $g_{\mu\nu}$ and
$D^\beta_{\mu\nu}|_\rho$ is the
non-Riemannian covariant derivative of the tensor $D^\beta_{\mu\nu}$
with respect to the full connection $\Gamma^\beta_{\mu\nu}$.

Continuing with
\begin{equation}
g_{\mu\nu}|_\rho=2g_{\mu\nu}\phi_\rho
\end{equation}
and the derived expression
\begin{equation}
\phi_\tau|_\rho=(\phi^\tau g_{\tau\mu})|_\rho=
\phi^\tau|_\rho g_{\mu\tau}+2\phi_\mu \phi_\rho \label{socd}
\end{equation}
used to eliminate terms in $\phi^\tau|_\rho$, there results
\begin{eqnarray}
R^\beta_{\cdot\mu\nu\rho}=Q^\beta_{\cdot\mu\nu\rho}
-\delta^\beta_\rho\phi_\nu|_\rho+\delta^\beta_\rho\phi_\rho|_\nu
-\delta^\beta_\nu\phi_{\mu\rho}-\delta^\beta_\rho\phi_{\mu\nu}
+g_{\mu\nu}g^{\lambda\beta}\phi_\lambda|_\rho
-g_{\mu\rho}g^{\lambda\beta}\phi_\lambda|_\nu \nonumber \\
-g_{\mu\nu}\phi^\beta\phi_\rho
+g_{\mu\rho}\phi^\beta\phi_\nu
+g_{\mu\nu}\delta^\beta_\rho \phi_\tau \phi^\tau
-g_{\mu\rho}\delta^\beta_\nu \phi_\tau \phi^\tau
-\delta^\beta_\rho \phi_\mu \phi_\nu
+\delta^\beta_\nu \phi_\mu \phi_\rho
\end{eqnarray}

The contraction with respect to first and third indices is
\begin{equation}
R^\nu_{\cdot \mu\nu\rho} \equiv R_{\mu\rho}=Q_{\mu\rho}
-3\phi_\mu|_\rho+ \phi_\rho|_\mu-g_{\mu\rho}\phi_\lambda|^\lambda
+2\phi_\mu \phi_\rho-2g_{\mu\rho} \phi^\tau \phi_\tau
\end{equation}
and contracting with the undotted inverse, $g^{\mu\rho}$, gives
\begin{equation}R^\nu_{\mu\nu\rho}g^{\mu\rho}=Q(g_{\mu\nu})
-6\phi_\lambda|^\lambda-6\phi_\lambda\phi^\lambda
\end{equation}

Reduction of the Weyl derivative can be made by
\begin{equation}
\phi_\lambda|_\nu g^{\nu\mu}
=\phi_{\mu ; \nu}-2\phi_\lambda\phi^\lambda
\end{equation}
which comes from equations~(\ref{socd}),~(\ref{nrcc}), and~(\ref{wegu})
and gives
\begin{equation}
R^\nu_{\mu\nu\rho} g^{\mu\rho}=
Q(g_{\mu\nu})-6\phi^\lambda_{\, ; \lambda}+6\phi^\lambda\phi_\lambda.
\label{gkge}
\end{equation}
The semicolon denotes the Riemannian covariant
derivative using the undotted christoffel symbols.
It is necessary to eliminate the undotted metric in favor of the
observers dotted metric and the conformal ratio $\lambda$.   The
fundamental quantity for the electromagnetic field is $\phi_\mu$ with
the index lowered.  The index of $\phi_\mu$ in equation~(\ref{gkge})
has been raised with the undotted metric.  The conventional
physical quantity is actually
\begin{equation}
\phi_\mu=\lambda\phi^\rho\dot g_{\mu\rho}
\end{equation}

The curvature scalar $Q(\dot g_{\mu\nu})$
of the dotted metric can be related to
the curvature scalar $Q(g_{\mu\nu})$ of the undotted metric by a
calculation
entirely analogous to those already done.  The complete expression becomes
\begin{eqnarray}
-6m^2=R_{\mu\nu} \dot g^{\mu\nu}=\lambda R_{\mu\nu}g^{\mu\nu}=
Q(\dot g_{\mu\nu})
+{3 \over \lambda \sqrt{- \dot g}}{\partial \over \partial x^\beta}
\left(\sqrt{-\dot g} \, \dot g^{\beta\alpha}
{\partial \lambda \over \partial x^\alpha} \right) \nonumber \\
-{3 \over 2 \lambda^2} \dot g^{\alpha\beta}
{\partial \lambda \over \partial x^\alpha}
{\partial \lambda \over \partial x^\beta}
-{6 \over \sqrt{- \dot g}}{\partial \over \partial x^\beta}
(\sqrt{-g} \, \phi^\beta)+6\lambda\phi_\beta\phi^\beta
\end{eqnarray}
Since $\phi_\beta$ is pure imaginary in this gauge, the linear term
separates and after substituting for $\phi_\beta$ with  $A_\beta$,
there results two real equations:
\begin{equation}
{{\partial \over \partial x^\beta}
(\sqrt{-\dot g} \, \dot g^{\beta\alpha} \lambda  A_\alpha)=0}. \label{gcpc}
\end{equation}
and
\begin{equation}
m^2-e^2 \dot g^{\mu\nu} A_\mu A_\nu=
{Q(\dot g_{\mu\nu})}
-{1 \over 2 \sqrt{-\dot g} \, \lambda}
{\partial \over \partial x^\beta} \left(
\sqrt{-\dot g} \, \dot g^{\beta \alpha}
{\partial \lambda  \over
\partial x^\alpha}\right) \\
+{\dot g^{\alpha\beta} \over 4 \lambda^2}  \,
{\partial \lambda \over \partial x^\beta}
{\partial \lambda \over \partial x^\alpha} \label{wkgr}
\end{equation}

The first of these expresses the conservation of the quantum probability
current as a congruence  of trajectories.
In the second equation, the left side includes all classical terms
from the Hamilton-Jacobi
equation for a general relativistic particle in an electromagnetic
field.  This can be seen by reversing the gauge transformations
and identifying the wave function with the action by
$\psi=\exp(iS)$.
The classical limit corresponds to neglecting derivatives
of the magnitude of the wave function.  The terms on the right side represent
curvature corrections due to either the Riemannian
curvature of the observer's space or the quantum curvature derived from the
conformal ratio $\lambda$.  The conformal contributions to the curvature
create the expected quantum effects.

Note that equations~(\ref{gcpc}) and~(\ref{wkgr})
are entirely real and do not contain any explicit reference to
the wave function $\psi$.
This form is useful because many previous five dimensional
theories~\cite{r3a}
are known that have two real fields $A_\mu$ and $g_{\mu\nu}$.
An attempt can be made to quantize some of
these by the substitution of the fixed gauge fields.
This process often seems to generate quantum terms and interpretations.
In other cases it is completely unworkable.
However, an approach of this type to first quantization avoids the serious
problems that have been discussed~\cite{r18}.

\section{CONFORMAL FACTORS}
\label{cff}

It is now essential to understand how the conformal factors developed
in (QG) can be used to form a field equation.  Only the most elementary
arguments are available because the standard methods that use lagrangians
fail.  Moreover the conformal factors do not have any established
interpretations.  The characteristic equations that contains these
factors are studied and then interpreted according to physical principles or
experiments.

It is by way of the quantum field equation that the mass must be introduced.
This external constant characterizes wave aspects of particle motion.
The introduction of the associated length scales must be compatible
with the scale sizes of other physical effects.
Boundaries, slits and shutters are scaled according to the masses
of the particles of which they are made.
Electromagnetic and gravitational source currents are scale sensitive
and require particles at measured positions.   The extrinsic
nature of mass is even suggested by particle creation.

It is known that free space
electromagnetic theory is invariant under conformal gauge change and
can contain no intrinsic scale size.
Scale dependence can enter only through the source terms.
Thus any distance dependency found in
electrodynamics must derive ultimately from the intrinsic
scale sizes of quantum source currents and must be related to the masses of
the source particles.
If the electric source equations can be combined with the quantum field
equations in a geometrical theory, the electric interaction constant becomes
geometrical and should have a geometrical interpretation.  This allows
the conversion of $e^2/mc$ to $e^2/\hbar c$.  It
may or may not be calculable, but one can anticipate that some theories will
have favored values.

Complex wave functions are used in the accepted description of quantum
mechanics~\cite{r72}.  They can be used legitimately in a geometrical
theory in so far as the phase and amplitude are combined into a linear
wave equation.  And while the quantum geodesics are completely real,
it seems reasonable to allow the use of complex
quantities in the search for fundamental invariants~\cite{r73,r52}.
The success of this approach is not assured, but if it happens that
such equations are identified, rearrangements to a
real theory should be possible.

Following (QG), source currents manifest their effects on
the metric through the conformal factors.
In constructing a theory of interaction, the three factors $\omega,\lambda,
\chi$ must depend on either the test
particle wave function $\psi_1$ or on one or
more source wave functions $\psi_2 \cdots \psi_N$.
It is by no means obvious which
combinations of fields $\psi_1 \cdots \psi_N$ are important in giving
values to each of $\omega, \lambda$, and $\chi$.  Various
geometrical quantities must be evaluated and compared with
known quantum equations.  Since none of these factors
affect the congruence of a specific quantum state, the
physical meaning must be inferred from the effect that they have
on the field equations.  The motion, as described by a particular
congruence, may be explained
by different combinations of source currents.
A simple form for interactions should be obtained by
allowing the source currents to affect the five metric
through the conformal factors.

A number of conformally covariant theories have used the
factors $\omega$ and $\chi$,  in addition to $\lambda$.
These quantities should all be identified with the
physics of interaction.  Applied as
contact transformations, the effect is to change
the five-gauge of $\gamma_{mn}$, the one-gauge of
the fifth dimension $\tau$ or the four-gauge of $g_{\mu\nu}$.
This includes, implicitly, the integrating factor
(or dis-integrating factor) of the vector potential.

A preliminary investigation $\omega$
shows that it may generate five-covariant interactions.  If,
following reference~\cite{r74}, the metric is conformally transformed,
so that the new Riemann tensor is
\begin{equation}
\overline \gamma_{mn}=e^{2\sigma}\gamma_{mn},
\end{equation}
then the new curvature is
\begin{eqnarray}
\overline \Theta_{mnab}= \nonumber \\
e^{2\sigma}
\left[\Theta_{mnab}
+\gamma_{ma}\sigma_{nb} +\gamma_{nb}\sigma_{ma} -\gamma_{mb}\sigma_{na}
-\gamma_{na}\sigma_{mb} +(\gamma_{ma}\gamma_{nb}-\gamma_{mb}\gamma_{na})
\gamma^{lt} \sigma_{,l} \sigma_{,t} \right]
\end{eqnarray}
and the Ricci tensor ${\overline \Theta_{mn}}$ is given by
\begin{equation}
{\overline \Theta_{mn}}=\Theta_{mn}+
(n-2) (\sigma_{;mn}-\sigma_{,m} \sigma_{,n})+
\gamma_{mn}[\gamma^{ab}\sigma_{;ab} +(n-2)\gamma^{ab}\sigma_{,a}\sigma_{,b}].
\end{equation}

If the lowest order dependence, perhaps as expressed in a local
euclidean region, is exponential, possibly of the form
$e^{i(\kappa x-\omega t+m\tau)}$, the lowest order
linear terms in velocity occur in the $(5\mu)$ positions and the quadratic
terms in the $(\mu\nu)$ positions.  The additivity is correct for the
source terms of classical equations.   A product of external
factors $e^{2\sigma_1} e^{2\sigma_2}$ will have linear additivity
with $(5\mu)$ terms of the form $\sigma_{1\mu} + \sigma_{2\mu}$ .
This is appropriate
for electromagnetic source currents.  The quadratic additivity of the
$\mu\nu$ terms will be of the
form $(\sigma_{1\mu}+\sigma_{2\mu})^2$ which is appropriate for
gravitational effects.  This suggests that, at least in the classical limit,
$\sigma_i$ must be part of the source current.  Probably in the
quantum limit, it must also be some part of the quantum source current.
This is apparently a
primitive type of interaction.  The internal conformal factors
may also participate but may reduce the invariance to four dimensions.

The association of  these factors with real source currents is still
under study.   A full understanding should give a
quantum version of the Maxwell-Einstein equations.
A prerequisite to such a  derivation will require an understanding
of how to interpret the coherent quantum terms geometrically.
Because, for this formalism, there is no classical foundation,
the problem of quantum understanding must be addressed first.  The
source currents are postponed and the quantum field equations
are taken up instead.

\section{CONFORMAL WAVES}
\label{cfw}

It seems reasonable to attempt to construct
a quantum equation in five space from an assumption about geometrical
invariants~\cite{r50}.  Since the coordinate $\tau$ is not
accessible, there
is no direct way to observe the actual system $(x^\mu,\tau)$.
Nevertheless, there
is  some indication of the importance of this formal coordinate.
It is that the collection of { \it non-inertial }
systems that transform among themselves by a cut transformation remain
physically equivalent until $\tau$ becomes a universal coordinate.
Otherwise stated, defining a fixed $\tau$ coordinate specifies a
particular gauge and the associated motion for each particle.
This specification represents
the beginning of a systemwide inertial structure. A similar idea has
been identified by Schouten~\cite{r51a}.

It is in practice the ratio of rest masses, that can be measured by
quantum diffraction.  One should therefore demand a theory for which
individual particles have constant rest mass ratios at points taken
where trajectories intersect.
Given any two distinct particles, $A$ and $B$, whose trajectories intersect at
a set of points.
$N_1,N_2, \cdots , N_k$, then there must be a way to
arrange the theory so that they  have equal effective mass ratios.
\begin{equation}
\left({M_A \over M_B}\right)_1=
\left({M_A \over M_B}\right)_2= \cdots =
\left({M_A \over M_B}\right)_k
\end{equation}
Such unvarying mass ratios of discreet particles measured quantum mechanically
is taken as an experimental fact.
To reproduce this observation with an extrinsic definition of mass,
the properties of each of the two particles must be referred
to a common geometrical system.
The neutral observer must be able to choose the rest
masses of particles so that they will not vary over space-time.
Otherwise the clocks cannot be
systematically calibrated. In so far as
the neutral observer can uniformly calibrate a system
of quantum clocks, the absolute mass of a particle can become a viable
global concept.

The concepts of mass and quantum time come together consistently if
the particle fields have a proper time dependence that appears always through
the product combination $m\tau$.
Now while the $\tau$ coordinate must be universal,
the fields observed in neutral space-time must also be $\tau$ independent.
For five dimensional invariance, field equations containing $\tau$
derivatives
are unavoidable.  These derivatives must operate on some residual $\tau$
dependence  which is then replaced by factors of $m$.  This departs from
Klein's assumption of the charge as an eigenvalue~\cite{r75} and
does not support a sequence of quantized values for the charge.

Qualitatively, one expects a Klein-Gordon equation
to appear as
\begin{equation}
\left( {\partial \over \partial t ^2}-
{\partial \over \partial x ^2}-
{\partial \over \partial y ^2}-
{\partial \over \partial z ^2}\right)\Psi=-m^2\Psi \equiv
{\partial \over \partial \tau ^2}\Psi
\end{equation}
with
\begin{equation}
\Psi=e^{im\tau}\psi(x^\mu)
\end{equation}
where the factor $e^{im\tau}$ applies only in a coordinate system
that has a fixed preferred alignment
to the observers' neutral frame.  The introduction of the mass reduces
the five dimensional dispersion free form and creates quantum
dispersion~\cite{r74a}.  This reduction
is essential and ultimately produces classical inertia.
Once this is done, the cut
transformations as an arbitrary coordinate transformation must
be relinquished as it will cause contact variations
of physical parameters.

In this way, the constancy of the mass ratios at each point,
can be assured from the requirement of
using a single fifth coordinate axis for all particles.
The universality of the five space also is inferred.
By using a real particle for the construction of clocks, and by choosing
a fixed numerical value of the mass, a uniform quantum clock is
generated.  At this point, coordinate transformations that affect the scale
size of the fifth coordinate axis
correspond to a continuous rescaling of all masses
(rather than individual masses) over space-time.
The quantum behavior references the fifth dimension,
which must be assumed common.

{}From this construction in five dimensions
it is seen that the mass spectrum consists of a single value.
A more complicated spectrum, might come from other geometries.
The elementary particles have distinct
masses but they are also all distinguished by additional
interactions.  Perhaps a geometrical theory that includes such
fundamental interactions will produce a more realistic mass spectrum.
Since the mass appears here as a property that is  extrinsic
to the particle, calculation of the mass from a field equation is
possible in principle.

The conformal waves can now be calculated for other cases.
An equivalent structure should occur in five dimensions. Let
a metric in $n$ dimensional space be of the form
$\omega \eta_{ma}$ where $\eta_{ma}$ is a unit diagonal tensor and $\omega$
is a manifold variable conformal factor.  The contribution of $\omega$ to
the curvature scalar is calculated directly.   It must be
linear in the second derivatives of $\omega$ and quadratic in the
first derivatives of $\omega$.  The numerical factors depend on the
number of dimensions $n$.

\begin{equation}
R=(n-1){1 \over \omega^2}
{\partial^2\omega \over \partial x^a \partial x_a} +
{(n-1)(n-6) \over 4} {1 \over \omega^3}
{\partial \omega \over \partial x^a}
{\partial \omega \over \partial x_a} \label {cnfw}
\end{equation}

A transformation of the form $\omega=\psi^p$ can be used to eliminate
the terms quadratic in the first derivative.
\begin{equation}
R \psi^p={(n-1)p}\left\{{1 \over \psi}
{\partial^2 \psi \over \partial x^2} + \left[
{(n-6) \over 4}p+(p-1)\right]{1 \over \psi^2}
{\partial \psi \over \partial x^\alpha}
{\partial \psi \over \partial x^\alpha} \right\} \label{gtlwe}
\end{equation}
Taking the term in brackets as zero gives
\begin{equation}
p={4 \over n-2}
\end{equation}
If $R=0$ be chosen as a fundamental invariant equation, a linear wave
equation for the conformal factor is generated except for the cases $n=1$ and
$n=2$.  The extended Weyl theory corresponds to $n=4,p=2$ with the
additional condition that $R\psi^p$ is constant.  This works physically
as long as a properly scaled base metric is available to form the
modified curvature scalar that absorbs the nonlinear factor
$\lambda=\psi^p=|\psi|^2$.  This is
automatically accomplished if quantum objects, transported from
region to region are used to make local
measurements.
In the five dimensional case $p=4/3$ and a term $R\psi^{4/3}$ is not linear.
The scalar $R$ must be chosen zero in agreement with the concept of null
five vectors.  The remaining terms give a wave equation of the expected form.
In almost any number of dimensions, conformal variations can be
used to generate a linear wave equation.

The conformal  waves of different dimensionality
may relate to each other nonlinearly.
A derived invariant linear equation for one value of $n$ will not
necessarily be linear as observed in a different number of
dimensions.
Because the fifth coordinate is not directly observable from space-time,
the internal conformal factor $\lambda$, (and indirectly $\chi$)
might also participate in various types of wave behavior.
With both factors involved together, it is necessary to
consider curvature scalars for a metric of the form
\begin{equation}
\pmatrix{\lambda\omega & & & & \cr & \lambda\omega & & & \cr
& & \lambda\omega & & \cr & & & \lambda\omega & \cr & & & & \omega \cr}
\label{cwmt}
\end{equation}
Here both $\lambda$ and $\omega$
are possibly dependent on five coordinates.

The curvature derivation
is carried out with all diagonal terms formally positive.  The
result for signature
$(1,-1,-1,-1,-1)$, can be found by appropriate sign changes.
Let $g_{mn}$ be diagonal but with possibly different values for
each element.  A long but elementary calculation gives
the Riemann tensor as
\begin{eqnarray}
 R_{abcd}=
{1 \over 2}(-g_{aa,bc}\delta_{ae}+g_{bb,ac}\delta_{be}
+g_{aa,bd}\delta_{ac}-g_{bb,ad}\delta_{bc}) \nonumber \\
+{\delta_{ad} \over 4}[g^{aa}g_{aa,b}g_{aa,c}+g^{bb}g_{aa,b}g_{bb,c}+
g^{cc}g_{aa,c}g_{cc,b}] \nonumber \\
-{\delta_{bd} \over 4}[g^{aa}g_{aa,c}g_{bb,a}+g^{bb}g_{bb,a}g_{bb,c}+
g^{cc}g_{cc,a}g_{bb,c}] \nonumber \\
-{\delta_{ac} \over 4}[g^{aa}g_{aa,d}g_{aa,b}+g^{bb}g_{aa,b}g_{bb,d}+
g^{dd}g_{aa,d}g_{dd,b}] \nonumber \\
+{\delta_{bc} \over 4}[g^{aa}g_{aa,d}g_{bb,a}+g^{bb}g_{bb,a}g_{bb,d}+
g^{dd}g_{dd,a}g_{bb,d}] \nonumber \\
-{1 \over 4}(\delta_{ad}\delta_{bc}-\delta_{ac}\delta_{bd})
\sum_m g_{aa,m}g_{bb,m}g^{mm}
\end{eqnarray}
in which all sums are written explicitly.

The Ricci tensor, given by the first contraction with the inverse metric
$g^{ac}=\delta^{ac}/g_{aa}$, is
\begin{eqnarray}
R_{bd}=
{1 \over 2}\left[\delta_{bd}\sum_m g^{mm} g_{bb,mm}
+\sum_m g^{mm} g_{mm,bd}
-g^{dd}g_{dd,bd}-g^{bb}g_{bb,bd}
\right] \nonumber  \\
+{1 \over 2}[(g^{dd})^2g_{dd,b}g_{dd,d}+(g^{bb})^2g_{bb,e}g_{bb,b}
+g^{dd}g^{bb}g_{dd,b}g_{bb,d}]   \nonumber \\
-{1 \over 4}  \sum_m \left[(g^{mm})^2g_{mm,d}g_{mm,b}+
g^{mm}g^{bb}g_{mm,b}g_{bb,d}+g^{mm}g^{dd}g_{mm,d}g_{dd,b}\right]  \nonumber \\
-{\delta_{bd} \over 2}\sum_m \left[
(g^{mm})^2 g_{mm,m}g_{bb,m}+g^{mm}g^{bb}g_{bb,m}g_{bb,m}\right] \nonumber  \\
+{\delta_{bd} \over 4} \sum_{m,n}g_{mm,n}g_{bb,n}g^{mm}g^{nn}
\end{eqnarray}
again with all sums explicit.
The scalar, given by another contraction with the inverse metric is
\begin{eqnarray}
R={3 \over 2}\sum_m(g^{mm})^3(g_{mm,m})^2
-\sum_m (g^{mm})^2g_{mm,mm}+\sum_{m,n}g^{mm}g^{nn}g_{mm,nn}
 \nonumber  \\
-\sum_{m,n}\left[(g^{mm})^2g^{nn}g_{mm,m}g_{nn,m}
+{3 \over 4}(g^{mm})^2g^{nn}g_{mm,n}g_{mm,n}\right] \nonumber  \\
+{1 \over 4}\sum_{m,n,p}g_{mm,p}g_{nn,p}g^{mm}g^{nn}g^{pp}
\end{eqnarray}
with all sums explicit.

This can be used to evaluate the diagonal case of equation~(\ref{cwmt})
by choosing
$g_{\mu\mu}=\omega\lambda,$ and $g_{55}=\omega$ and rearranging, using
$x^a=x^\alpha$ for $a=\alpha=1,2,3,4$ and $x^5=\tau$.
\begin{eqnarray}
R={3 \over \lambda^2 \omega}
{\partial^2 \lambda \over (\partial x^\beta)^2}
-{3 \over 2\lambda^3 \omega}
\left( { \partial \lambda \over \partial x^\beta } \right)^2
+{4 \over \lambda \omega^2}
{\partial^2 \omega \over (\partial x^\beta)^2}
-{1 \over \lambda \omega^3}
\left( {\partial \omega \over \partial x^\beta} \right)^2
+{4 \over \omega^2} {\partial^2 \omega \over \partial \tau^2}
-{1 \over \omega^3} \left( {\partial \omega \over \partial \tau } \right)^2
\nonumber \\
+{4 \over \lambda^2 \omega^2}
{\partial \omega \over \partial x^\beta}
{\partial \lambda \over \partial x^\beta}
+{8 \over \lambda \omega^2} {\partial \omega \over \partial \tau }
{\partial \lambda \over \partial \tau}
+{4 \over \lambda \omega} {\partial^2 \lambda \over \partial \tau^2} \
+{1 \over \lambda^2 \omega}
\left( {\partial \lambda \over \partial \tau} \right)^2
\label{loreq}
\end{eqnarray}

It is easy to see that $\lambda$ and $\omega$ both
have wave properties which can be coupled to each other by an
imposed  constraint on the curvature  scalar.
It is suggestive to loosly
associate $\omega$ and $\lambda$ with wave functions.
{}From this expression, the coupling terms are
of the form $A_\mu J^\mu$.
Considering first just the $\lambda$ wave, and removing the coupling
by setting $\omega=1$, the expression for the curvature scalar becomes
\begin{equation}
R=-{3 \over \lambda^2}{\partial^2 \lambda \over (\partial x^\mu)^2}
+{3 \over 2 \lambda^3}\left({\partial \lambda \over \partial x^\mu}\right)^2
-{4 \over \lambda}{\partial^2 \lambda \over \partial \tau^2}
-{1 \over \lambda^2}\left({\partial \lambda \over \partial \tau }\right)^2
\label{confw}
\end{equation}
Because the first two terms will reduce to the form $\Box \psi /\psi$ after
the substitution $\lambda=\psi^2$,  a four dimensional covariant equation is
possible for the four dimensional conformal waves that are internal to the
five dimensional metric.

Alternativly, the same sort of calculation of scalar curvature dependence
on $\omega$ with $\lambda=1$ gives the result predicted by
equation~(\ref{cnfw}).
Both types of waves may be important to define a form of
unified interaction.   The following sections consider these
two cases with some care to evaluate gravitational and
electromagnetic effects.

\section{INTERNAL KLEIN-GORDON EQUATION}
\label{inkg}

Consider first the four dimensional conformal factor.  Because
in equation~(\ref{confw}),
the terms in $\tau$ do not have the same degree of homogeneity in $\lambda$
as the terms in $x^\mu$,
the $\tau$ dependence cannot be used to generate a mass term.
Apparently, this structure corresponds more to the Weyl theory in which the
mass is introduced as an external constant.
The resulting equation is not a five dimensional invariant and the mass
must scale against the observer's metric.
As with the Weyl theory, the observers' metric must
be set up so that a quantum object, perhaps an hydrogen atom, is
spherically symmetric and of constant size. Relative to this
particular system, conformal changes must be executed with care
and with consideration for the gauge system.
Measurements of the gauge of the observed metric, $\dot g_{\mu\nu}$
in separated regions of space-time must match up when
expanded to overlap.

The gauge factor $\lambda$ is a neutral space four-scalar but might also
be a function of $x^5=\tau$.  This dependence is presumed to be of the
form $e^{imb \tau}$.  Any other
proper time dependencies might show effects that would be
explicitly observable by measuring quantum probability densities.
Most physical interpretations support the
interpretation of $\lambda$ as a function only of $x^\mu$.
This transformations does leave the trajectories invariant.  With this factor
included, it is the most general factor
that is allowed for the conditions imposed.

The calculation of curvature, including electromagnetic and gravitational
effects, is somewhat involved.
By transforming the coordinate system at a fixed but arbitrary point $P$,
it is possible to reduce $\gamma_{mn}$ to a locally pseudo-euclidean
system with diagonal$(1,-1,-1,-1,-1)$ and with derivatives
$\gamma_{mn,a}=0$.   This is the local co-moving frame that,
as required by five space equivalence, removes all interactions.
The electromagnetic field becomes zero
since then $\A_{\mu,\nu}=0$ and
$H_{\mu\nu} \equiv \A_{\mu ,\nu}-\A_{\nu,\mu } =0$.  This transformation
would require the observed macroscopic fields $\dot g_{\mu\nu}$ and $A_\mu$
to be explicitly $\tau$ dependent.  A milder
transformation that looks like it is four dimensional to the
observer is more useful.

Let the four-space coordinates $x^\mu$ be transformed at
a fixed point $P$ so that
${\partial \dot g_{\mu\nu} / \partial x^\beta} = 0$ making the
neutral space locally geodesic.  Now perform a cut transformation
\begin{equation}
\tau^\prime=\tau + \Pi(x^\mu) \label{ctintkg}
\end{equation}
and choose $\Pi$ so that
$\A_\mu^\prime \Bigl|_P =0$ and
$(\A_{\nu,\mu}^\prime + \A_{\mu,\nu}^\prime) \Bigl|_P =0$.
This value can be calculated explicitly if
\begin{equation}
\A_\mu^\prime =\A_\mu + \Pi_\mu \label{ngtc1}
\end{equation}
and
\begin{equation}
\A_{\mu,\nu}^\prime =\A_{\mu,\nu} + \Pi_{\mu\nu}. \label{ngtc2}
\end{equation}
Assuming a polynomial expansion gives
\begin{equation}
\Pi = -\A_\mu\Bigl|_P x^\mu
-{1 \over 4} ( \A_{\mu,\nu} + \A_{\nu,\mu})\Bigl|_P x^\mu x^\nu
\label{gtrf}
\end{equation}
The first condition~(\ref{ngtc1}) simplifies the metric at $P$ so that
it is diagonal while the second is a local Killing condition for
$\A_\mu^\prime$.  Of course $H_{\mu\nu}^\prime=H_{\mu\nu}$ is gauge
invariant  and four covariant
as it should be.  The reverse transformation to~(\ref{gtrf}) must be performed
in order to return to the coordinate system of the original metric.

The calculation of the Riemann tensor
can be carried out at point $P$ and the full tensor
regenerated later.  The result is unwieldy in some cases.
This requires values of $\gamma^{mn}$, the Christoffel
symbols $[m,np]$, and their derivatives $[m,np],q$.  Evaluations at point
$P$ are simpler because many quantities are zero. In particular,
\begin{eqnarray}
\dot g_{\mu\nu , \sigma}\Bigl|_P=\Sigma_{\mu\nu} \Bigl|_P =
\A_\mu \Bigl|_P = \A_{\mu,5}=\dot g_{\mu \nu,5} = 0 \\
\gamma_{mn}\Bigl|_P =
\pmatrix{ \lambda \dot g_{\mu \nu} & 0 \cr 0 & -1 \cr } \\
\end{eqnarray}

Where $g_{\mu \nu} = \lambda \dot g_{\mu \nu}$ ,
$H_{\mu \nu} = \A_{\mu , \nu}- \A_{\nu , \mu}$, and
$\Sigma_{\mu \nu} = \A_{\mu , \nu}+ \A_{\nu , \mu}$.
The Christoffel symbols are
\begin{eqnarray}
2[\alpha, \beta \gamma]= g_{\alpha\beta,\gamma}
+g_{\gamma\alpha,\beta}
-g_{\beta\gamma,\alpha}
-(\A_\alpha \A_\beta)_{,\gamma}
-(\A_\gamma \A_\alpha)_{,\beta}
+(\A_\beta \A_\gamma)_{,\alpha} \\
2[5,\beta \gamma]= \Sigma_{\beta\gamma}  - g_{\beta \gamma,5} \\
2[\alpha,5 \gamma]=g_{\alpha \gamma,5} + H_{\alpha \gamma} \\
2[\alpha,55]=2[5,\alpha 5]=2[5,55]=0 .
\end{eqnarray}
These lead to the calculated values,
\begin{eqnarray}
2[\alpha, \beta \gamma]\Bigl|_P = g_{\alpha\beta,\gamma}
+g_{\gamma\alpha,\beta}
-g_{\beta\gamma,\alpha}=
\dot g_{\alpha\beta} \lambda_{,\gamma}
+\dot g_{\gamma\alpha} \lambda_{,\beta}
-\dot g_{\beta\gamma} \lambda_{,\alpha} \\
2[5,\beta \gamma]\Bigl|_P= - g_{\beta \gamma,5}
= -\dot g_{\beta \gamma} \lambda_{,5} \\
2[\alpha,5 \gamma]\Bigl|_P = g_{\alpha \gamma,5} + H_{\alpha \gamma}
= g_{\alpha \gamma}\lambda_{,5} + H_{\alpha \gamma}
\end{eqnarray}
and
\begin{eqnarray}
2\left. {\mu \brace \beta \gamma} \right| _P =
\delta^\mu_\gamma {\lambda _{,\beta} \over \lambda}
+\delta^\mu_\beta {\lambda _{,\gamma} \over \lambda}
-\dot g_{\beta \gamma} \dot g^{\alpha \mu} {\lambda_{,\alpha} \over \lambda}
-{(\A^\mu \A_\beta)_{,\gamma} \over \lambda}
-{(\A^\mu \A_\gamma)_{,\beta} \over \lambda}
+{(\A_\beta \A_\gamma)_{,\alpha} \dot g^{\alpha \mu} \over  \lambda} \\
2 \left. {5 \brace \beta \gamma}\right|_P =
-\sigma_{\beta \gamma} + \dot g_{\beta \gamma} \lambda_{,5} \\
2 \left. { \alpha \brace 5 \gamma}\right|_P =
{\delta_\gamma^\alpha \lambda_{,5} \over \lambda}
+{ H^\alpha_{\cdot\gamma} \over \lambda} .
\end{eqnarray}

And for the derivatives at point $P$
\begin{eqnarray}
2[\alpha,\beta \gamma],\epsilon = g_{\alpha \beta ,\gamma \epsilon}
+g_{\alpha \gamma,\beta \epsilon}
-g_{\beta \gamma,\alpha \epsilon}
-H_{\alpha \gamma} \A_{\beta,\epsilon}
-H_{\alpha \beta} \A_{\gamma,\epsilon}= \nonumber \\
\dot g_{\alpha \beta,\gamma \epsilon}
+\dot g_{\alpha \gamma,\beta \epsilon}
-\dot g_{\beta \gamma,\alpha \epsilon}
+\dot g_{\alpha \beta} \lambda _{,\gamma \epsilon}
+\dot g_{\alpha \gamma}\lambda_{,\beta \epsilon}
-\dot g_{\beta \gamma} \lambda_{,\alpha \epsilon}
-H_{\alpha \gamma} \A_{\beta,\epsilon}
-H_{\alpha \beta} \A_{\gamma,\epsilon} \\
2[5,\beta \gamma],\epsilon = \Sigma_{\beta \gamma,\epsilon}
-g_{\beta \gamma},_{5 \epsilon} \\
2[\alpha,5 \gamma],\epsilon = \dot g_{\alpha \gamma ,5 \epsilon}
+H_{\alpha \gamma,\epsilon} \\
2[\alpha,\beta \gamma],5 = \dot g_{\alpha \beta} \lambda_{,\gamma 5}
+\dot g_{\alpha \gamma} \lambda_{,\beta 5}
-\dot g_{\beta \gamma} \lambda_{,\alpha 5} \\
2[5,\beta \gamma],5 = -\dot g_{\beta \gamma} \lambda_{,55} \\
2[\alpha,5 \gamma],5 = \dot g_{\alpha \gamma} \lambda_{,55}.
\end{eqnarray}

For the Riemann tensor calculation, terms can be grouped by whether they
are quantum mechanical $Q_{abcd}$ with factors of
$\lambda$, electromagnetic $E_{abcd}$ with factors
of $H_{\mu\nu}$ or gravitational $R_{abcd}$
with factors of $\dot g_{\mu\nu,\alpha \beta}$.
Depending on the type of theory, these terms may transform into each other.
In five dimensions the full tensor can be written
\begin{equation}
\Theta_{abcd} \equiv Q_{abcd}+R_{abcd}+E_{abcd}=
{}[a,bc],d-[a,bd],c+\gamma^{tu}[t,bd][u,ac]-\gamma^{tu}[t,bc][u,ad]
\end{equation}
Three classes of terms can be considered and calculated separately
depending on whether there is no index equal to $5$, one index equal to $5$
or two indices equal to $5$.
For the first case,
\begin{equation}
R_{\alpha \beta \gamma \epsilon}= {1 \over 2}
\left[\dot g_{\alpha \gamma,\beta \epsilon}
-\dot g_{\beta \gamma,\alpha \epsilon}
-\dot g_{\alpha \epsilon,\beta \gamma}+
\dot g_{\beta \epsilon,\alpha \gamma} \right],
\end{equation}
\begin{eqnarray}
Q_{\alpha \beta \gamma \epsilon} =
{1 \over 2} \left[ \dot g _{\alpha \gamma} \lambda_{,\beta \epsilon}
-\dot g _{\beta \gamma} \lambda_{,\alpha \epsilon}
-\dot g _{\alpha \epsilon} \lambda_{,\beta \gamma}
+\dot g _{\beta \epsilon} \lambda_{,\alpha \gamma} \right] \nonumber \\
+{1 \over 4} \left[g_{\tau \beta} \lambda_{,\epsilon}
+g_{\tau \epsilon} \lambda_{,\beta}
-g_{\beta \epsilon} \lambda_{,\tau} \right] g^{\tau \mu}
\left[g_{\mu \alpha} \lambda_{,\gamma} +g_{\mu \gamma} \lambda_{,\alpha}
-g_{\alpha \gamma} \lambda_{,\mu} \right] \nonumber \\
-{1 \over 4} \left[g_{\tau \beta} \lambda_{,\gamma }
+g_{\tau \gamma} \lambda_{,\beta}
-g_{\beta \gamma} \lambda_{,\tau} \right] g^{\tau \mu}
\left[g_{\mu \alpha} \lambda_{,\epsilon} +g_{\mu \epsilon} \lambda ,_\alpha
-g_{\alpha \epsilon} \lambda_{,\mu} \right] \nonumber \\
-{1 \over 4} g_{\beta \epsilon} g_{\alpha \gamma} (\lambda_{,5})^2
+{1 \over 4} g_{\beta \gamma} g_{\alpha \epsilon} (\lambda_{,5})^2
\end{eqnarray}
and
\begin{equation}
E_{\alpha \beta \gamma \epsilon} = \A_{\alpha,\epsilon} \A_{\beta,\gamma}
-\A_{\alpha,\gamma} \A_{\beta,\epsilon}
-2\A_{\alpha,\beta} \A_{\gamma,\epsilon}
\end{equation}
For the terms containing one $5$
\begin{equation}
R_{\alpha \beta \gamma 5}=0,
\end{equation}
\begin{equation}
Q_{\alpha \beta \gamma 5} ={1 \over 2}
( \dot g_{\alpha \gamma} \lambda_{,\beta 5}
- \dot g_{\beta \gamma} \lambda_{,\alpha 5})
+{1 \over 2 \lambda } (g_{\beta \gamma} \lambda_{,5} \lambda_{,\alpha}
-g_{\alpha \gamma} \lambda_{,5} \lambda_{,\beta}),
\end{equation}
and
\begin{equation}
E_{\alpha \beta \gamma 5} =
{1 \over 2}(H_{\alpha \beta} \lambda_{,\gamma} -H_{\alpha \beta,\gamma})
+{1 \over 4}(H_{\gamma \beta} \lambda_{,\alpha}
-H_{\gamma \alpha} \lambda_{,\beta}
+g_{\beta \gamma} \lambda_,^{\cdot \tau} H_{\tau \alpha}
-g_{\alpha \gamma} \lambda_,^{\cdot \tau} H_{\tau \beta}).
\end{equation}

For terms containing two $5$'s,
\begin{equation}
R_{\alpha 5 \beta 5} =0,
\end{equation}
\begin{equation}
Q_{\alpha 5 \beta 5}=
{1 \over 2} \dot g_{\alpha \beta} \lambda_{,55}
-{1 \over 4\lambda} \dot g_{\alpha \beta} \lambda_{,5} \lambda_{,5}
\end{equation}
and
\begin{equation}
E_{\alpha 5 \beta 5}= -{1 \over 4\lambda} H^\tau_\alpha H_{\tau \beta}.
\end{equation}
All other four index terms are zero.

The Ricci tensor
$\Theta_{ac} = \Theta_{abce}\gamma^{be}$
can be split up the same way,
\begin{equation}
R_{\alpha \gamma}={1 \over 2 \lambda}
[\dot g_{\alpha \gamma,\beta \epsilon} \dot g^{\beta \epsilon}
-2\dot g_{\beta \gamma,\alpha \epsilon}\dot g^{\beta \epsilon}
+\dot g_{\beta \epsilon,\alpha \gamma}\dot g^{\beta \epsilon}],
\end{equation}
\begin{equation}
Q_{\alpha \gamma}={1 \over 2 \lambda}
(\dot g_{\alpha \gamma} \lambda_{\mu \tau} \dot g^{\mu \tau}
+2 \lambda_{,\alpha \gamma})
-{3 \over 2 \lambda^2} \lambda_{,\alpha} \lambda_{,\gamma}
+{1 \over 2} \dot g_{\alpha \gamma}
\left( -{\lambda_{,5} \lambda_{,5} \over \lambda } +\lambda_{,55} \right)
\end{equation}
and
\begin{equation}
E_{\alpha \gamma} = -{1 \over 2 \lambda}
H_{\alpha \mu} H_{\gamma \tau} \dot g^{\mu \tau}.
\end{equation}
Also
\begin{equation}
Q_{\beta 5} = {3 \over 2} \left( {\lambda_{,\beta 5} \over \lambda}
-{\lambda_{,\beta} \lambda_{,5} \over \lambda^2} \right)
\end{equation}
and
\begin{equation}
 E_{\beta 5}= -{1 \over 2\lambda} H_{\alpha \beta,\gamma}
\dot g^{\alpha \gamma}.
\end{equation}
Then
\begin{equation}
R_{55} =0
\end{equation}
\begin{equation}
Q_{55} ={2 \lambda_{,55}  \over \lambda}
- {(\lambda_{,5})^2 \over \lambda^2}
\end{equation}
and
\begin{equation}
E_{55} =-{1 \over 4 \lambda^2} H_{\alpha \beta} H_{\mu \tau}
\dot g^{\alpha \mu} g^{\beta \tau}
\end{equation}
Finally, the five scalar curvature can be calculated
\begin{equation}
R = {\dot R \over \lambda},
\end{equation}
\begin{equation}
Q = {3 \over \lambda^2}\lambda_{,\mu \tau} \dot g^{\mu \tau}
-{3 \over 2 \lambda^2} \lambda_{,\mu} \lambda,_\tau \dot g^{\mu \tau}
-{4 \over \lambda} \lambda_{,55}
-{(\lambda_{,5})^2  \over \lambda^2}
\end{equation}
and
\begin{equation}
E = -{1 \over 4 \lambda^2} H_{\alpha \beta} H_{\mu \tau}
\dot g^{\alpha \mu} \dot g^{\beta \tau}
\end{equation}
These quantities are basic four-space
invariants of the metric $\gamma_{ab}$.
The term Q has the form of a Klein-Gordon equation, including correct
electromagnetic effects when retransformed
back to the original coordinate system using the inverse of
equations~(\ref{ctintkg}) and~(\ref{gtrf}).
Even with the electromagnetic and gravitational terms intact,
this result  is entirely similar to the Weyl theory.  The
five dimensional effects take the place of the non-Riemannian terms.

There are two characteristics of this
construction which make it useful.
Presumably any interaction
which can be described by a congruence has  a
Weyl like form.  And
as expected, the calculation of a four dimensional conformal variation within
a five dimensional space generates a reasonable and expected result.
The more important point is that whatever
structure quantum mechanics has, if it is to be described in a four
dimensional space-time, it must satisfy certain covariance
properties.  This derivation defines some of the combinations of
field quantities that can be used while allowing for the known,
accepted, covariance of space-time measurements.

\section{INVARIANT KLEIN-GORDON EQUATION}
\label{ikg}

The five dimensional conformal factor $\omega$ generates a different aspect
of a quantum system.
If, as suggested by the condition
of nullity, R is set equal to zero, then equation~(\ref{gtlwe}) for
n=5 is apparently also equivalent to the
Klein-Gordon equation.   The terms in equation~(\ref{loreq})
$\partial^2 \omega / (\partial x^\mu)^2 $ and
$(\partial \omega / \partial x^\mu)^2$ combine, as suggested into a linear
equation of the proper form and in this case the terms in
$\partial^2 \omega / \partial \tau^2$ and
$(\partial \omega / \partial \tau)^2$ can be included.
An exponential $\tau$ dependence for $\omega$ is acceptable and provides
a means to introduce the mass.

Following
section~(\ref{cfw}), the curvature scalar, $\Theta$, should be set to zero.
Conformal waves in $\gamma_{mn}$ are defined by
\begin{equation}
\gamma_{mn}=\omega \dot \gamma_{mn} .
\end{equation}
Where here $\dot \gamma_{mn}$ is a representation of local
electromagnetic and gravitational effects.
The derivation follows reference~\cite{r74} with $\omega=e^{2\sigma}$.
The scalar curvature depends on $\omega$ and the  untransformed metric
$\dot \gamma_{jk}$ according to
\begin{equation}
\Theta={1 \over \omega}\left[\dot \Theta +
{4 \over \sqrt{\dot \gamma} \, }{\partial \over \partial x^j}
\left({\sqrt{\dot \gamma} \, \dot \gamma^{jk} \over \omega}
{\partial \omega \over \partial x^k} \right)
+{3\dot \gamma^{jk}\over \omega^2}
{\partial \omega \over \partial x^j}
{\partial \omega \over \partial x^k}\right]
\end{equation}

As suggested by the calculation of section~(\ref{cfw}),
a rearrangement of terms that are powers of $\omega$ gives with $\Theta=0$
and $\Psi=\omega^{3/4}$,
\begin{equation}
0= \dot \Theta +
{16 \over 3 \Psi \sqrt{\dot \gamma} \, }{\partial \over \partial x^j}
\left(\sqrt{\dot \gamma} \, \dot \gamma^{jk}
{\partial \Psi \over \partial x^k} \right).
\end{equation}
This can be compared to Klein's paper where it is shown that the appropriate
equation is
\begin{equation}
{1 \over \sqrt{\dot \gamma}}{\partial \over \partial x^m}
\left(\sqrt{\dot \gamma} \, \dot \gamma^{mn}
{\partial \Psi \over \partial x^n}\right) =0 \label{kkg}
\end{equation}
with
\begin{equation}
\Psi=\psi(x^\mu)e^{im\tau}.
\end{equation}

They are the same except for the term
$\dot \Theta = R-{1\over 4} H_{\mu\nu} H_{\beta\lambda} \dot g^{\mu\beta}
\dot g^{\nu\lambda}$ which is small unless extremely high fields are present.
It appears additive to $m^2$.  In the unusual
situation where $\dot \Theta$ may be large, the effect is to cause
the particle to propagate as if it had a different rest mass.
The gravitational part of $\dot \Theta$ is equal
to the local gravitational scalar curvature.  As discussed in~\cite{r19}, it
is very small on the atomic scale and it is conventional
to choose the laboratory coordinates so that it is zero. There may
be some extraterrestrial circumstances where this term is important.
The electromagnetic component is very small, of order $\sim 10^{-40}$
for the Klein normalization of $\dot g_{\mu\nu}$.  It is an electrodynamic
correction to $\Theta$ that can become important if the electromagnetic
potential changes significantly over a Planck length.
Such a process could be
related to particle creation or other violent effects at high
energy~\cite{r74b}.
Since the term is additive to the mass,  particles in very strong
electromagnetic fields will propagate as if their effective rest mass
is altered.  In this case, because of the sign, the effect is to
reduce the rest mass in regions near a localized point charge,
allowing for the possibility of solutions having space
like trajectories for very short distances.  The fundamental concept
of space can be retained but the notion of time like motion fails.
The uncertainty principle, being derived from the properties
of this differential equation, can also fail in regions of high field.
In an extended theory, this type of effect may be important for describing
the motion of light particles, such as electrons, inside the nucleus.
The particle mass for the Klein normalization of $\dot g_{\mu\nu}$
must be a large value numerically.
To what extent this affects the possible relationship between the
Planck clock
and the quantum clock depends on the choice of the remaining
field equations.

At this point, many of the questions of inertia discussed in (QG)
are resolved.  The mass, as an external constant, reproduces
quantum diffraction and classical inertia.  It is also apparent that
a quantum clock can be standardized relative to the assumed
mass cofactor in the product $m\tau$.  The five dimensional space thus
combines the concepts of Mach, Newton and Einstein.
It is absolute, as Newton would have assumed.  It has no overall intrinsic
inertial structure as Mach would have assumed.  It provides a mathematical
substrate for four dimensional covariance as Einstein would have insisted.

As is well known, measurements of the gauge of the observed
metric, $\dot g_{\mu\nu}$
in separated regions of space-time match up when the separated
space-time regions are expanded to overlap. This is
automatically accomplished if quantum objects, transported from
region to region are used to make local
measurements during determination of $\dot g_{\mu\nu}$.
The quantum behavior references the fifth dimension,
which must be assumed common to the entire space-time,
and guarantees the length scales.

Because this equation is based on an invariant, it can be used to
calculate the results of quantum mechanics in arbitrary gravitational fields.
In the classical limit, the Klein-Gordon equation reduces to the
Hamilton-Jacobi equation, and the macroscopic manifestations of this field
equation will match the known properties of a classical charged particle.
The gravitational mass
remains positive and unchanged for antiparticles, which should be in agreement
with experiment~\cite{r76}.  This limit seems also to agree with the question
of coordinate conditions in reference~\cite{r77}.

Once a local metric is established, quantum effects can be derived.
For instance, in a general case one could start with $j_\mu$ and
$T_{\alpha \beta}$ as classical  source currents
and get in the classical limit $A_\mu$ and $g_{\mu\nu}$.
In other cases, the calculation of fields is not part of the experiment and
they are determined by direct local measurements.  The assumption of
a universal value for the metric $\dot g_{\mu\nu}$ can be the basis for
a calculation of isolated quantum effects.
Either way, the terms in $\dot g_{\mu\nu}$ and the vector potential provide
a starting point and give a fundamental five dimensional tensor
$\gamma_{mn}$.  This method might not work if essential quantum
interactions occur between the source and test particle.  The
result should hold for a separated quantum experiment with external
classical interactions.

The theory can be applied to diffraction of neutrons~\cite{r78} or
atoms~\cite{r79} in situations where only semiclassical theories have
previously been used.   As the nuclear phase shifts are
not part of geometrical theories so far developed,  the phenomenological
values need to be inserted. The theoretical situation is better
for atomic diffraction where, excepting spin, the capabilities
of this analysis are adequate. See also reference~\cite{r79a}.

Earthbound experiments may not be useful for testing some of the predictions.
Large gravitational fields in which gravitational non-linearities are
significant would be required to observe differences from the classical
relativistic calculations.  Significant experiments are difficult to find
and an evaluation may be inconclusive until a more complete theory is known.
The best type of test might involve more complicated theories that include
nuclear or particle properties.

A set of field equations that apply to coherent sources is needed.
The simple substitution of coherent for incoherent terms cannot be
correct because the coherent effects are not always additive in the classical
limit.  Second order terms in the source fields
are required to reproduce the known dependence of the gravitational potential
on the quantum state.
The effect is required by the classical principle of equivalence applied to
internal state changes of the sources~\cite{r77a}.
The gravitational potential of a superimposed beam of electrons
and protons will depend on whether they have combined to form hydrogen.
The effective mass density depends on the quantum state and hence {\it
cannot be
calculated classically}. The second order terms analogous to those
in the Schr\"odinger or  Klein-Gordon equation are required
to extract the correction. Some purely gravitational theories~\cite{r27}
have suggestive second order factors.

The problem of coherence is more apparent for
photon correlation experiments.  If the detectors are
treated as classical source currents, the
correlation cannot be predicted because there are no absorbing quantum
states that can be used to identify the properties
of the participating photons. The correlations
are in principle still present because the advanced potentials
must be used to generate the radiative forces which cause the absorption
and emission.
The photon correlations however, cannot be observed without accounting for
the inherently non-linear quantum effects.   For this reason,
a correct mathematical theory of interaction
must be sophisticated enough to include quantum electrodynamics in that
limit.

Finally, the question of the establishment of a coordinate system from
quantum measurements can be addressed.
Specifically, the null geodesics of (QG)
reduce to the classical geodesics that form the epistemological
basis of covariant measurements.
They are conceptually sufficient for the construction of a
general coordinate system.  By choosing experiments in which $\hbar/m$ is
effectively small compared to other distances, the simplified motion of
classical
particles results directly .   This is a critically important process because
it indicates how a quantum general relativistic theory can reduce to
classical general relativity.  Moreover, it generates the accepted scheme for
constructing the coordinate system of the neutral observer.  The probability
density trajectories match the classical solutions of the Hamilton-Jacobi
equation  to lowest order in $\hbar$.  In the classical limit, the fixed
gauge condition for the vector potential can be relaxed.  The magnitude of
the wave function is effectively constant and the phase or action can
transform with the gauge of the vector potential.  The arbitrary motion
associated with diffraction and interference is gone because of the
assumed simplification of $\psi$.
Thus, from the quantum geodesics and the concept of null five-curvature,
comes a metaphysical basis for the operational demonstration of
space-time metrics in arbitrary coordinate systems.
\footnote{
The classical limit can also be studied as a limit of wave packet motion.
This also works since without the higher order terms, the
packets becomes non-dispersive and follow the Hamilton-Jacobi trajectories.
The wave packet case is, perhaps more realistic while the DeBroglie-Bohm
like trajectory formalism may be more relevant to some situations.}

\section{DISCUSSION}
\label{disc}

Some of the probable consequences of the geometrical
approach are worth noting.
Studies are continuing with regard to interaction theory
and constants.  Of immediate interest is some set of quantum
Einstein-Maxwell equations. It appears that this will use a Klein like
expression for the gravitational constant and that the fine structure
constant will be an internal geometric ratio.
It is a complicated problem because the
interaction mechanism between five dimensional representations of
quantum particles needs to be observed from neutral space-time.
At least three metrical structures are needed.
In addition, the known allowed solutions for
two particles without background interactions are
restricted~\cite{r80}.
Additional distant interacting particles, that participate in
the quantum-gravitational-electrodynamic boundary conditions,
may be important, at least to generate the equivalent of free fields.
The problem is related to the arguments of Renninger~\cite{r81}.
It is necessary to include all possible emitting and absorbing particles
in the system if normal radiative behavior is to be expected.

Because of the analysis of conformal waves, it is
easy to see that the number of theories that can generate a conformal
interpretation for quantum effects is limited.  Probably in a five
dimensional theory, only the four and five dimensional waves are
realistic candidates.  Since the quantum structure must
be introduced integrally, other possibilities may not be available.

The mathematical differences between this theory and the conventional ones
may allow  new predictions for quantum
electrodynamics.  Few such effects are expected
to be observable in the laboratory.  For instance, one could
consider a photon correlation experiment in which the photon
energies or polarizations are changed by a gravitational field
between the emitter and the detector.  One would expect an appropriate
correlation to be present.
The laboratory optical correlation experiments do not differentiate
between quantum and geometrical theories.  Further experiments
of this type seem unnecessary.  Prediction of
the correlations seems to be accomplished by
time symmetric potentials interacting nonlinearly with geometrical
quantum states.

A greater concern is the inhomogeneous quantum-Maxwell equations that
are required  for the calculation
of the electrodynamic field.  These appear to be similar to the classical
Maxwell equations except for the substitution of quantum source currents.
This part of the structure
is not yet a unified geometrical reality.  A definitive resolution
of the photon correlation experiments must await a
complete set of source equations.  A five dimensional theory
would also reasonably include quantum gravitational effects.
In the meanwhile, it is necessary to argue physically
that the accepted quantum-Maxwell equations are adequate.
Never the less, it is easy to see that once
the form of the electromagnetic interactions is chosen to give
the correct results for common electromagnetic experiments, the
photon correlations will be predicted correctly.

Because any motion, possibly even non-inertial,
 can be described by a five-coordinate transformation,
it appears that an absolute global five space can
be chosen flat for some types of interaction theory.
A particular quantum-gravitational-electromagnetic state
$(\psi ,A_\mu , g_{\mu\nu})$ may  then be embedded in some sense.
The congruence of trajectories becomes part of a curvilinear coordinate
system.   This embedment is not the same as the usual general
relativistic problem of embedding~\cite{r82}
because only one quantum state
is embedded at a time.  The conventional approach seeks to embed the entire
classical dynamics at each point.  Thus it must include all classically
allowed initial conditions.  The quantum case, as approached by the
geometrical method, is much simpler because
at each point the congruence has only one direction and magnitude.
Each velocity at each given point is part of a different physical problem.
For changed boundary or initial conditions,
new fields are chosen and a new embedding calculation is to
performed.  It is in part because of this simplification that
an {\it elementary} five dimensional theory of interaction may be possible.

It is expected that the external invariant Klein-Gordon equation
applies as a primitive constraint on the motion of the test
particle while the internal four dimensional equation
is to be applied to the source particles as a condition on the
description of the source current.
The condition that the source terms have the proper dependence on the
wave function means that the two interpretations of $\psi$, as
either an electric source current
or as a probability density can be identified with each other.
It is this equality
that allows the definition of the electric charge.  For
the primitive geometrical Maxwell's equations, the concept of stream
electricity must be used~\cite{r83}
which has the interpretation
here of mutually interpenetrating geometrical congruences.  The exact
relation of the internal and external Klein-Gordon equations
must be part of an interaction theory.

This relationship between currents and particle motion
forms a hidden variable theory of the crypto-deterministic type.
That is, while it is technically possible to describe particles as moving
points on trajectories, the experimental elucidation of those trajectories
is not possible.  For a true, physically detectable,
hidden variable theory, one would imagine that
the unity of the congruence is broken and that separate parts of it
could interact with different observers.  That this is not the case
is manifested by having the interacting fields depend on
source currents represented only as complete congruences
without identified specific trajectories.
Other forms of interaction could conceivably
destroy the particle quantum coherence.  There is no experimental
evidence for such a failure of this ``stream electricity''.
It is appropriate to assume that geometrical theories should have
congruence based interactions and remain crypto-deterministic.
This is only possible if one rejects the notion of a classical
basis for quantum theory.
Apprently, the semiclassical radiation theories produce incorrect
results because a
classical point particle is placed
on a calculated trajectory. The source of the field must
be the quantum current as a whole.

Thus the trajectories are not required to have a separate existence.
They may however be treated as a conceptual aid to the
visualization and calculation of congruences.   One may also choose to ignore
the crypto-determinism and maintain a version of wave-particle duality.
The establishment of quantum particle trajectories as
real ontological objects is not necessary.  Like most fields,
they have been invented for
 mathematical convenience and metaphysical comfort.

The difference between incoherent and coherent sources for electromagnetism
should be noted.  The experimental
inverse square law for point particles appears only in the
classical limit when the wave function has been sufficiently contracted.
It does not apply to a discrete quantum
particle traveling on the congruence.  The classical limit must come
when the congruence itself
is bundled into a small space.  In this limit each coherent
source term can be treated as one of a collection of incoherent particles.
The classical electrodynamics obtains.  It is this
inverse square law of a well localized packet, that is universally
accepted.  The kernel used
in quantum electrodynamics is quite different metaphysically and must
be applied only to the quantum current to determine the interaction of the
geometrical streams.  By making this distinction explicit,
problems of semi-classical radiation
theory~\cite{r84} are avoided.
The construction of a classical inverse square interaction must not be
supposed instantaneous, but is the sum of effects, propagated
at advanced or retarded times in accord with the requirements of
relativistic invariance.

One looks to Kaluza's theory to supply the quantum source terms that
are necessary.
It fails in detail because the coupling to the quantum density cannot
readily be specified in a theory that is constructed on a classical
basis.  The Kaluza results should be interpreted as a sort of classical
limit.   Unfortunately, the extended principle of equivalence plus the
concept of implicit quantization means that quantum mechanics cannot readily
be separated from this electrodynamics. There is no
obvious way to quantize Kaluza's theory for the same
reasons that there is no way to quantize general relativity.  One can
only hope to derive these as the limit of some higher construction.
The standard Einstein-Maxwell construction demonstrates some of the
same difficulties.  It has no source terms and it may not be possible to
give it five-covariant source terms except in the quantum case.
As derived, these equations are quantum suppressed and can at most apply
to a classical system.

Second quantization is not directly addressed here, but a few comments
should indicate the approach.  The complicated
formalism of equal time commutators is not needed.
These have the apparent function of introducing the
additional derivatives (to make second order field equations) in a
way that can keep them separated during multiparticle interactions.
Basically, the second order terms implied by the commutation relations
must each be identified with a
separate particle.  The field quantities of distinct particles
must always commute.  In a geometrical theory, differential
invariants generate these second order terms from the curvature calculations.
The construction of multiple metrics that is presented
in (QG) replaces the equal time commutation relations.   Each particle
has an associated metric, and from it, the second order terms
are generated from the invariant tensors.
The commutation between distinct particle fields is automatic.

If the source density can be integrated into such a theory, the
fine structure constant may be implicitly defined.  Curvature terms
suggest factors of the form
${(n_1 / 4 \pi n_2)}$ for small integers $n_1, n_2$.  Such a value
is approximately ten times larger than the
measurements.  Any final relation awaits a firmer theory.  A value that
agrees with experiment is probably not to be expected since
the issues of spin and weak or strong interactions are not addressed.
The process of renormalization may also be important, but this
construction takes a very different form in an
exact quantum-gravitational theory.  Any further consideration of these
matters is intended to be taken up at a later time.

While the literature on spin is extensive, there are several studies
that may be useful to mention.  The
problem has been discussed for both five dimensional theories~\cite{r85},
and for Weyl theories~\cite{r86}.
Any of these must yet be adapted to this approach with a fixed
gauge and geometrical interpretation.
The fundamentality of fermions and their universal description in
terms of complex four by four spinors suggests that the Dirac spinors,
five in number, may be used to define an extension to five dimensional
geometry.  It is anticipated that the vector congruence
should be replaced by a pentad (funfbein) congruence.  A better understanding
of spin in this context is required.
Because of the change in the nature of spinor representations beyond $n=5$,
further extension may be limited.
These issues may also be considered in more detail later.

The justification for associating the conformal machinery of a geometrical
theory with quantum mechanics is crucial to the validity of this approach.
It is an interesting hypothesis that
a wholly geometrical theory of quantum mechanics can be found.
So far in the literature, very little has been done, especially for theories
that do not derive
from a classical model.   The conformal factors are studied here because
they seem to be the simplest possibility and  demonstrate a
characteristic quantum behavior.

To evaluate this approach, it is reasonable to ask whether
the results of experiments are correctly explained and whether the
mathematical system is free of damaging inconsistencies.  These results
appear to be better than the standard formalism on both accounts.
They allow for the discussion
of quantum effects in the presence of gravitational fields and they
avoid the serious problems due to an inconsistent use of derivatives.
While there is much more to do, the possibility of further predictions of
physical importance is a realistic claim.

\section{SUMMARY}
\label{sum}

A five dimensional covariant theory is studied as a mechanism to discover
how quantum fields can be made
integral to a system based on classical differential
geometry.  The basic mechanism is to understand how internal and external
conformal factors might be related to the
source fields that can produce a given metric.

The non-Riemannian Weyl theories demonstrate how
the quantum structure can be interpreted.  This calculation,
repeated in fixed gauge form, produces a five dimensional derivation.
 A study of the conformal factors indicate that the wave behavior
of the five metric is related to the wave functions of the test and source
particles.  It becomes possible to view quantum field equations
as a set of conditions on the curvature tensors.

Some of the metaphysical problems of combining general relativity and
quantum mechanics are resolved.  Quantum mechanics has been given
a geometrical origin.  General relativity has a space time structure
that is accurately represented by the trajectories of a quantum particle
in the classical limit.  Both are to be developed from a more
fundamental and primitive geometry.  A primitive electrodynamics
is also present.  Reference to classical theory is avoided.

The introduction of the mass as an extrinsic quantum constant generates
a unified theory of inertia that reproduces classical
and quantum observations.
As an eigenvalue, it  produces the correct
mass dependence and the correct quantum field equation.  The fifth
dimension is necessary. The resulting interaction
theory appears to be reasonable if applied by using classical equivalence.
It is also possible to calculate quantum effects in gravitational
fields that are classically generated.    The geodesics can
be used to construct the perceived classical
coordinate system if combined with fundamental quantum clocks.

The combined field system predicts
diffraction and interference effects while ascribing the actual
motion to the null geodesics.  The Klein-Gordon equation
gains a term that may be important for high
field densities and for situations where field derivatives are
significant over particle scales.  Further studies are needed to establish
a mathematically closed system of equations.  The quantum
source currents for the calculation of the fields must yet be defined.
Some of the problems of doing this are discussed.

Beyond new understanding, the important result is that there seems to
be a way to combine gravitation with other interactions
into a covariant theory.  It is at least successful in that it shows that
the geometrical perspective can be extended to include the basic quantum
processes.

\vfill \eject
\narrowtext
\figure{Under a continuous conformal transformation in four dimensions, the
measured value of the probability density is modified by the local conformal
parameters.  The actual number of counts for a fixed region of space-time is
invariant because of the adjustment to the numerical volume.\label{f6} }
\end{document}